\documentclass{article}
\usepackage{ijcai20}

\usepackage{times}
\usepackage{soul}
\usepackage{url}
\usepackage[hidelinks]{hyperref}
\usepackage[utf8]{inputenc}
\usepackage[small]{caption}
\usepackage{graphicx}
\usepackage{amsmath}
\usepackage{amsthm}
\usepackage{booktabs}
\usepackage{algorithm}
\usepackage{bm}
\usepackage{pbox}
\usepackage{color}
\usepackage{wrapfig}
\usepackage{amssymb}
\usepackage{multirow}
\usepackage{algorithmicx}
\usepackage{algpseudocode}

\title{Inferring Degrees from Incomplete Networks and Nonlinear Dynamics}

\author{
Chunheng Jiang\and
Jianxi Gao\And
Malik Magdon-Ismail
\affiliations
Rensselaer Polytechnic Institute\\
110 8th Street, Troy, NY 12180\\
\emails
\{jiangc4,gaoj8\}@rpi.edu,
magdon@cs.rpi.edu}

\begin{document}
\maketitle

\begin{abstract}
Inferring topological characteristics of complex networks from observed data is critical to understand the dynamical behavior of networked systems, ranging from the Internet and the World Wide Web to biological networks and social networks. Prior studies usually focus on the structure-based estimation to infer network sizes, degree distributions, average degrees, and more.  Little effort attempted to estimate the specific degree of each vertex from a sampled induced graph, which prevents us from measuring the lethality of nodes in protein networks and influencers in social networks. The current approaches dramatically fail for a tiny sampled induced graph and require a specific sampling method and a large sample size. These approaches neglect information of the vertex state, representing the dynamical behavior of the networked system, such as the biomass of species or expression of a gene, which is useful for degree estimation. We fill this gap by developing a framework to infer individual vertex degrees with both the sampled topology and vertex states. We combine the mean-field theory with combinatorial optimization to learn vertex degrees. Experimental results on real networks with a variety of dynamics demonstrate that our framework can produce reliable degree estimates and dramatically improve existing link prediction methods by replacing the sampled degrees with our estimated degrees.
\end{abstract}

\section{Introduction}
Many networks, although large in sizes, are substantially incomplete ~\cite{barabasi2002evolution,szilagyi2005prediction,woodward2010ecological,chen2016seeing}  when compared with the real systems.  To learn the topological characteristics of a complex network, we often need powerful accessing of the vertices and edges of the whole system. However, the associated cost is prohibitive, especially for large-scale networks, e.g., social networks with millions or even billions of nodes. 
Therefore, we seek a reliable and cost-effective way to estimate graph characteristics from incomplete (sampled) data~\cite{ribeiro2012estimation,eden2018provable}.

The individual vertex degrees provide a cornerstone to have reliable estimations of many graph statistics, including network size~\cite{bunge1993estimating,stumpf2008estimating,katzir2011estimating,kurant2012graph}, average degree~\cite{goldreich2008approximating,dasgupta2014estimating} and motif counting~\cite{klusowski2018counting,kashtan2004efficient}. The vertex degree measures the lethality of a node in protein-protein interaction networks~\cite{jeong2001lethality}, which is usually unknown if the network is incomplete.  Consequently, estimating individual degree has its practical meaning in understanding the cellular function. Nevertheless, it is a challenging problem. Although many studies are concentrating on graph sampling techniques to infer some macroscopic properties of networks, such as the \textit{degree distribution} or the associated \textit{moments of degree distribution}~\cite{frank1980estimation,stumpf2005sampling,achlioptas2009bias,ribeiro2012estimation,zhang2015estimating}, very few work on estimating the \textit{individual vertex degrees}~\cite{ganguly2017estimation} so far. Ganguly and Kolaczyk~\shortcite{ganguly2017estimation} derived a class of estimators from a risk-theoretic perspective and provided analytical analysis for them. These estimators take full advantage of the local topology information in the sampled sub-network to infer the individual degrees of the sampled vertices. However, the wide variety in the network structure can make the attempt in vain if we do not have any knowledge of specific global network characteristics, such as species abundance in ecological networks, or gene expression in biological networks. However, such crucial information is available but was not used in the risk-minimization framework. 

\begin{figure}[htbp]
\centering
\includegraphics[width=0.45\textwidth]{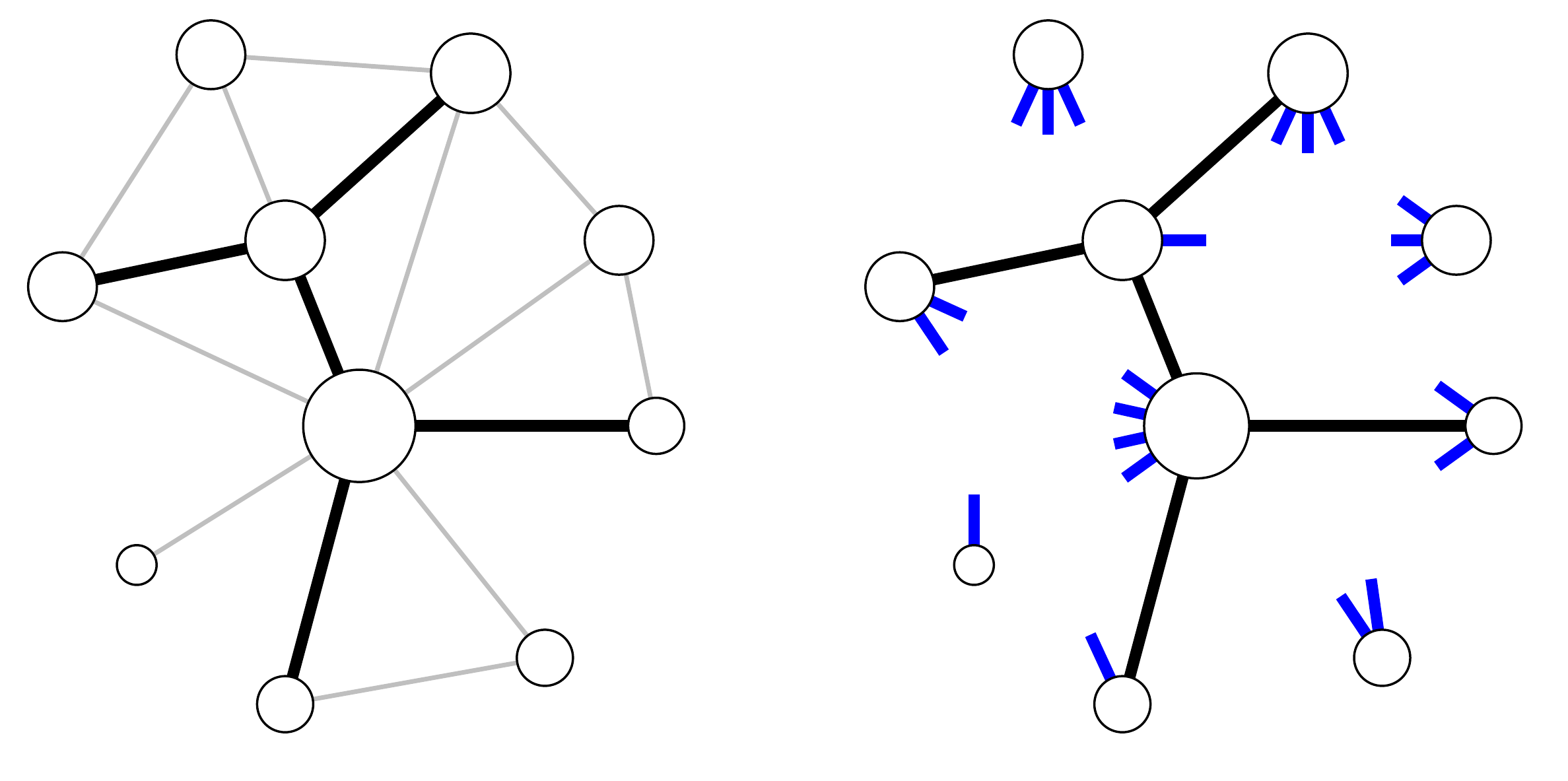}
\caption{Illustration of our algorithm. (Left), an incomplete network is
  observed (black edges) and the state values of the dynamics at each vertex
  is indicated by the
  size of the vertex. (Right) Our work uses the state values, dynamics
  plus the incomplete
network to infer the degrees at each vertex (blue stubs).}
\label{fig:illustration}
\end{figure}

In our framework, assuming that we are given a set of coupled nonlinear equations governs dynamics of the network, and the steady-states of all the nodes or a subset of nodes under the dynamics have been observed.
We propose to estimate the individual vertex degrees
from its steady-state without knowing any topology of the network, and we further improve the accuracy of the prediction if we know a tiny percentage of sampled graph regardless of sampling methods, as illustrated in Figure~\ref{fig:illustration}.
The motivation is that the steady-state of some process (e.g., an epidemic spreading, gene expression)
may be observable even though the full network is not observable. For example, in social networks, we may know who got sick, but we may do not know who interacts with whom; in biological networks, we may know the gene expression of all the genes of a patient, but we may not completely know the wiring diagram of them. It is possible to infer individual vertex degrees because the steady-state is the outcome of a complex interplay between the network's topology and the dynamics. 
Our framework does not rely on any prior knowledge of the sampling methods or the degree distribution. However, it offers excellent inference with as low as 10\% of known edges, and moderate level of noises in the measurements. 
Another advantage is its ability to recover individual degrees for all vertices with only partial topology, (in)accurate measurements of the state information, though the dynamical system may be misspecified. To demonstrate the impacts of sampling scheme and the measurement noise in steady-states, and how the local topology and the dynamics contribute to the overall accuracy, we conduct a series of empirical experiments on various real networks. Experimental results indicate that our approaches can accurately infer individual vertex degrees and very robust to the sampling scheme. The inference quality can be guaranteed as long as the measurements of the steady-state are not distorted too much.

\paragraph{Problem Setup.}
Let the implicit full network be $G=(V,E)$. It is only partially observable from a sub-network $G^{(s)}=(V^{(s)},E^{(s)})$ with $V^{(s)}\subset V$ and $E^{(s)}\subset E$. Let ${\bm \delta}=(\delta_1,\delta_2,\ldots,\delta_n)$ be the true individual vertex degrees, and $n=|V|$. To be convenient, bold letters hereinafter in mathematical formulas represent vectors.

We consider a general dynamics in which each vertex of $G$ evolves according to a self-driving force and a sum of interaction forces over the neighbors
\begin{equation}
\dot x_i = f(x_i) + \sum_{j\in V} A_{ij} g(x_i,x_j),\forall i.
\label{eq:dynsys_ode}
\end{equation}
Both $f(\cdot)$ and $g(\cdot,\cdot)$ are general and usually nonlinear, and the adjacency matrix $A\ge 0$ modulates the interactions between vertices. The coupled ordinary differential equations (ODEs)
which are based on the complete topology of $G$ drive the states of the
vertices to an equilibrium $\bm x^*$. Given the
complete network, this equilibrium can be approximated
efficiently using a  mean-field approximation derived
in~\cite{gao2016universal}
\begin{equation}
\dot x_i = f(x_i) + \delta_i g(x_i,x_{\rm eff}),\forall i.
\label{eq:mf_ode}
\end{equation}
This mean-field approximation interprets the impacts of a neighborhood using
an average impact determined by
$x_{\rm eff}$,
which is equivalent to a homogeneity assumption.
The total interactions of the neighbors of a vertex amounts to
multiplying this average interaction by $\delta_i$.
To compute the mean-field steady-states
$\bm{z}^*$
in \eqref{eq:mf_ode}, one  needs vertex degrees
$\delta_i$ and the effective network impact, $x_{\rm eff}$, which
can be computed given the complete network.
Our contribution is to show how to compute these steady-states given only
the incomplete network, thereby establishing a relationship between the
observed steady-states and the vertex degrees, from which we back-out the 
individual degrees. To fill in the details, we first describe the
mean-field approximation.

\section{The Effective Network Impact}
\label{sec:xeff}
\def\degin{\delta^{\rm in}}
\def\degout{\delta^{\rm out}}

Now consider a vertex $i$ and the interaction term
$\sum_j A_{ij} g(x_i,x_j)$ in \eqref{eq:dynsys_ode}, where 
$A_{ij}$ is the influence $j$ has on $i$. Similarly,
$i$ influences $j$ with a weight
$A_{ji}$. We define
the in-degree $\degin_i=\sum_j A_{ij}$ and the out-degree
$\degout_i=\sum_{j}A_{ji}$. Assuming $A_{ij}\ge 0$, the interaction can be written as 
\begin{equation}
\sum_j A_{ij} g(x_i,x_j) = \degin_i\frac{\sum_{j} A_{ij} g(x_i,x_j)}{\sum_{k} A_{ik}}.
\end{equation}
Here, the in-degrees $\bm \degin$
captures the idiosyncratic part, and the average $g(\cdot,\cdot)$ captures
the network effect.
The mean-field approximation is to replace
local averaging with global averaging, which approximates the
network-impact on a vertex as nearly homogeneous. Specifically, we have
\begin{equation}
\label{eq:MF_av}
\resizebox{.91\linewidth}{!}{$
\frac{\sum_{j} A_{ij} g(x_i,x_j)}{\sum_{k} A_{ik}}
\approx
\frac{\sum_{ij} A_{ij} g(x_i,x_j)}{\sum_{i k} A_{i k}}
=
\frac{{\bm 1}^T A g(x_i,{\bm x})}{{\bm 1}^T A {\bm 1}},
$}
\end{equation}
where the vector $g(x_i,\bm x)$ has 
the $j$th component $g(x_i,x_j)$. We define a linear operator
\begin{equation}
{\cal L}_A(\bm z)=\frac{{\bm 1}^T A }{{\bm 1}^T A {\bm 1}}{\bm z}
=\frac{{\bm \degout}\cdot \bm z}{{\bm \degout}\cdot {\bm 1}}
\end{equation}
to produce a weighted average of the entries in $\bm z$.
Then, in our mean-field approximation,
\begin{equation}
\dot x_i=f(x_i)+\degin_i {\cal L}_A[g(x_i,\bm{x})].
\label{eq:approx1}
\end{equation}
In the first order linear
approximation, we can take the ${\cal L}_A$-average inside $g$.
The average of external interactions is approximately the interaction with
its average.  That is
${\cal L}_A[g(x_i,\bm{x})]\approx g(x_i,{\cal L}_A(\bm{x}))$ and
\begin{equation}
\dot{x}_i=f(x_i)+\degin_i g(x_i,{\cal L}_A(\bm{x})),
\label{eq:approx2}
\end{equation}
where ${\cal L}_A(\bm x)$ is a global state. Let $x_\text{av}\triangleq{\cal L}_A(\bm x)$. Applying
${\cal L}_A$ to both sides of
\eqref{eq:approx2} gives
\begin{equation}
\dot{x}_\text{av}={\cal L}_A[f(\bm x)]+
{\cal L}_A[{\bm \delta}^{\rm in} g(\bm x,x_{\text{av}})].
\end{equation}
According to the extensive discussion and tests in Ref.~\cite{gao2016universal}, the
in-degrees $\bm\degin$ and the interaction with the
external $x_{\rm av}$ are roughly
uncorrelated, so the ${\cal L}_A$-average of the product is roughly
the product of
${\cal L}_A$-averages.
Thus, we have
${\cal L}_A[\bm\degin g(\bm x,x_{\text{av}})]
\approx{\cal L}_A(\bm\degin){\cal L}_A[g(\bm x,x_{\text{av}})]$.
Using the first order linear approximation, we take the ${\cal L}_A$-average
inside $f$ and $g$
\begin{equation}
\dot{x}_{\text{av}}
= f({\cal L}_A(\bm x))+ {\cal L}_A(\bm \degin) g({\cal L}_A(\bm x),x_{\text{av}}).
\end{equation}
Now we have
\begin{equation}
\dot{x}_{\text{av}}=
f(x_{\text{av}})+
\beta g(x_{\text{av}},x_{\text{av}}),\label{eq:dotxeff}
\end{equation}
where the resilience $\beta={\cal L}_A(\bm{\degin})$.
For undirected graphs, ${\bm \degin}={\bm \degout}$, so
$\beta=\sum_i\delta_i^2/\sum_i\delta_i =\langle{\bm\delta}^2\rangle/\langle{\bm\delta}\rangle$.
The steady-state of \eqref{eq:dotxeff} is the effective network impact, $x_{\rm eff}$.
Plugging it into
\eqref{eq:approx2} gives
an uncoupled ODE for $x_i$,
\begin{equation}
\dot{x}_i=f(x_i)+\degin_i g(x_i,x_{\rm eff}).
\end{equation}
In the mean-field
approximation, $g(x_i,x_j)$ in 
\eqref{eq:dynsys_ode} is replaced by an interaction with a mean-field
external world
$g(x_i,x_{\rm eff})$ and the number of
neighbors impacting $x_i$ is captured by $\degin_i$.
To approximately obtain the steady-states of
the system, one first solves the ODE in
\eqref{eq:dotxeff} to get $x_{\rm eff}$,
and then $n$ uncoupled ODEs at each vertex
to get $x_i$, which only depends on
$\degin_i$ if given $x_{\rm eff}$. The method works well because the mean-field approximations
only need to hold \emph{at the steady-state}. Hence, we can recover the steady-state for any vertex (for
example the sampled vertices) from accurate estimates
 $\degin_i$ ($\delta_i$ in the undirected case) and $\beta$. 

\section{Our Approaches}
Our approaches need the observed steady-states $\bm x^*$ of nodes to estimate the individual node degrees $\bm\delta$, which can be improved if we also know the partial topology of the network. We consider the observed $\bm x^*$ as the ground truths and aim to search an optimal combination of degrees such that the predicted steady-states $\bm z^*$ of the approximated system \eqref{eq:mf_ode} are close to $\bm x^*$. In this sense, the general techniques underlying our work can be formulated as an optimization problem:
\begin{align}
\label{eq:est_deg_prob}
\min_{\hat{\bm\delta}} & ~~~~{\rm Err}(\bm x^*,\bm z^*)\nonumber \\
s.t. &~~~~ \hat\delta_i = \delta_i^{(s)}+d_i,\nonumber \\
&~~~~  \hat\beta = \langle \hat{\bm\delta}^2\rangle/\langle \hat{\bm\delta}\rangle,\\
&~~~~  f(\hat x_{\rm eff}) + \hat\beta g(\hat x_{\rm eff},\hat x_{\rm eff})=0,\nonumber \\
&~~~~  f(z_i^*)+\sum_j A_{ij}^{(s)} g(z_i^*,x_j^*) + d_i g(z_i^*,\hat x_{\rm eff})=0,\nonumber 
\end{align}
where $\delta_i^{(s)}$ is the degree of the sampled vertex $i$ in the incident/induced subgraph, $d_i$ is the estimated missing degree for $i$ (see more details in Section~\ref{sec:enhance}).

First, we propose the {\bf ZeroTopo} algorithm to estimate the degree of each node without knowing any topology of the network, i.e., $A^{(s)}$ has only zero elements. Then, we design an efficient 2-step procedure -- {\bf TopoPlus} and {\bf Round} -- to improve the accuracy of degree prediction by introducing the observed topology. 

\begin{itemize}
\item {\bf ZeroTopo} estimates ${\bm \delta}$ and $x_{\rm eff}$ with only the observed steady states ${\bm x}^*$ and with ZERO known topology.
\item {\bf TopoPlus} estimates ${\bm\delta}$ and ${\bm d}$ with both the partially observed topology and ${\bm x}^*$. Thus, $\hat{\bm\delta}={\bm\delta^{(s)}}+{\bm d}$, meaning that the estimated degree of a node is equal to the sum of its sampled degree and the missing degree.
\item {\bf Round} further optimizes \eqref{eq:est_deg_prob} for \emph{integer} missing degrees $\hat{\bm d}$, and the final estimate becomes $\hat{\bm \delta}^*={\bm\delta^{(s)}}+\hat{\bm d}$.
\end{itemize}

\begin{algorithm}[H]
\caption{ZeroTopo}
\begin{algorithmic}[1]
\Require $\bm x^*$ 
\State{Let $t=0$ and $x_{\rm{eff}}^{(0)}=\langle {\bm x}^*\rangle$}
\State{Solve $\hat\delta_i^{(1)}$ from \eqref{eq:mf_ode} with $x_{\rm eff}^{(0)}$}
\Repeat
\State{Compute $\hat\beta^{(t)}=\langle {\hat{\bm\delta}}^{(t)^2}\rangle/\langle \hat{\bm\delta}^{(t)}\rangle$}
\State{Solve $x_{\rm{eff}}^{(t)}$ from \eqref{eq:dotxeff} with $\hat\beta^{(t)}$}
\State{Solve $\hat\delta_i^{(t+1)}$ from \eqref{eq:mf_ode} with ${\bm x^*}$ and $x_{\rm eff}^{(t)}$}
\Until{$\hat\beta^{(t)}$ and $x_{\rm eff}^{(t)}$ do not change}
\Ensure $\hat {\bm \delta}=\lfloor\hat {\bm \delta}^{(t)}\rceil$
\end{algorithmic}
\label{alg:initialization}
\end{algorithm}

\subsection{ZeroTopo}
{\bf ZeroTopo} refers to the estimation of vertex degree based on observed steady-states $\bm x^*$ and the dynamics $f(\cdot)$ and $g(\cdot,\cdot)$, without any known topology information, as shown in Algorithm~\eqref{alg:initialization}. The average steady-state is set as the initial value of the effective state $x_{\rm eff}=\langle {\bm x}^*\rangle$ when {\bf ZeroTopo} warms up (line 1). Based on the mean-field approximation, we can repeatedly estimate the resilience index $\beta$ from estimated degrees and update the effective state such that the entire system is self-consistent. The algorithm is simple but very powerful to recover all individual degrees simultaneously. 

\begin{algorithm}[H]
\caption{TopoPlus}
\label{alg:enhancement}
\begin{algorithmic}[1]
\Require $\bm x^*,\bm \delta^{(s)}, A^{(s)}$ 
\State{Let $t=0$ and $\hat{\bm \delta}^{(t)}={\bm \delta}^{(s)}$}
\Repeat
\State{Compute $\hat\beta^{(t)}=\langle {\hat{\bm\delta}}^{(t)^2}\rangle/\langle \hat{\bm\delta}^{(t)}\rangle$}
\State{Solve $x_{\rm{eff}}^{(t)}$ from \eqref{eq:dotxeff} with $\hat\beta^{(t)}$}
\State{Solve $d_i^{(t)}$ from \eqref{eq:missed_degree} with $A^{(s)}$, ${\bm x^*}$ and $x_{\rm eff}^{(t)}$}
\State{Update $\hat\delta_i^{(t+1)}\leftarrow \delta_i^{(s)}+d_i^{(t)}$}
\Until{$\hat\beta^{(t)}$ and $x_{\rm eff}^{(t)}$ do not change}
\Ensure $\hat {\bm \delta}=\lfloor\hat {\bm \delta}^{(t)}\rceil$, ${\bm d}=\lfloor {\bm d}^{(t)}\rceil$ and $\hat x_{\rm eff}=x_{\rm eff}^{(t)}$
\end{algorithmic}
\end{algorithm}

\subsection{TopoPlus}\label{sec:enhance}
In the {\bf ZeroTopo} algorithm, we assume that we have no prior information about the topology of the network. In this section, we aim to enhance the accuracy of the degree prediction when we observe a subgraph $G^{(s)}$. Note that the {\bf TopoPlus} algorithm only requires the topology of the observed subgraph regardless of how one samples it. We incorporate the adjacent matrix $A^{(s)}$ into \eqref{eq:mf_ode} to identify the missing degree $d_i$ in the discarded local topology. Hence, we get an enhanced mean-field approximation

\begin{equation}
\dot x_i=f(x_i)+\sum_j A_{ij}^{(s)} g(x_i,x_j^*)+d_i g(x_i,x_{\rm eff}),
\label{eq:enhanced_mf_ode}
\end{equation}
with the steady-states $\bm x^*$ of the observed vertices, and the effective network impact $x_{\rm eff}$. Once $x_{\rm eff}$ is found, it is easy to derive $d_i$, that is
\begin{equation}
d_i=-\big[f(x_i^*)+\sum_j A_{ij}^{(s)} g(x_i^*,x_j^*)\big]/g(x_i^*,x_{\rm eff}).
\label{eq:missed_degree}
\end{equation}
The term $d_i$ is a compensation to $\delta_i^{(s)}$, i.e., $\hat\delta_i=\delta_i^{(s)} + d_i$. The idea lies in the mean-field approximation \eqref{eq:mf_ode} that $\sum_j A_{ij} g(x_i^*,x_j^*)\approx \delta_i g(x_i^*,x_{\rm eff})$. We do not have \emph{all} connections $\delta_i$ associated with vertex $i$, but we have \emph{some}, $\delta_i^{(s)}\le\delta_i$.  Based on the same approximation \eqref{eq:mf_ode}, we know $\sum_j A_{ij}^{(s)} g(x_i^*,x_j^*)\approx \delta_i^{(s)} g(x_i^*,x_{\rm eff})$, motivating us to search the missing $d_i=\delta_i-
\delta_i^{(s)}$ from \eqref{eq:enhanced_mf_ode} and \eqref{eq:missed_degree}.
Note that for vertices not in the sampled vertices, $\delta_i^{(s)}=0,\forall i\not\in V^{(s)}$. As a result, the enhanced approximation \eqref{eq:enhanced_mf_ode} is equivalent to \eqref{eq:mf_ode}, then computing $d_i$ in \eqref{eq:missed_degree} is exactly solving $\hat\delta_i$ from \eqref{eq:mf_ode} in {\bf ZeroTopo}.
{\bf TopoPlus} is formulated in Algorithm~\eqref{alg:enhancement}.

\begin{algorithm}[H]
\caption{Round}
\begin{algorithmic}[1]
\Require ${\bm x}^*,{\bm\delta}^{(s)},{\bm d},\hat x_{\rm eff}$
\Ensure $\hat{\bm \delta}^*={\bm\delta}^{(s)}+{\bm d}$
\Repeat
\State{Let $z_i^*(d_i)$ be a solution to}
\Statex{~~~~~~~~~~~~~~~$\dot z_i=f(z_i)+\sum_{j} A_{ij}^{(s)} g(z_i,x_j^*)+d_i g(z_i,\hat x_{\rm{eff}})$}
\State{Let $\varepsilon_i(d_i)={\rm Err}\{z_i^*(d_i),x_i^*\}$}
\State{Compute $Q_i^+=\varepsilon_i(d_i+1)-\varepsilon_i(d_i)$, }
\Statex{~~~~~~~~~~~~~~~~~~~~~~$Q_i^-=\varepsilon_i(d_i-1)-\varepsilon_i(d_i)$,$\forall i$}
\Statex{\Comment{\textit{isolated nodes are exclusive in calculating $Q^-$}}}
\State{Obtain nodes $I^+$ with $Q_i^+\le 0$, $I^-$ with $Q_i^- \le 0$}
\State{Let $I=I^+ \cap I^-$, $I^+\leftarrow I^+-I$ and $I^-\leftarrow I^--I$}
\For{each $(i,j)\in (I^+,I^-)$}
\State{$d_i\leftarrow d_i+1$, $d_j\leftarrow d_j-1$}
\EndFor
\For{each $i\in I$}
\If{$|Q_i^+|>|Q_i^-|$}
\State{$d_i\leftarrow d_i+1$}
\ElsIf{$|Q_i^+|<|Q_i^-|$}
\State{$d_i\leftarrow d_i-1$}
\EndIf
\EndFor
\State{Let $\hat{\bm\delta}^*={\bm\delta}^{(s)}+{\bm d}$, and calculate $\hat\beta=\langle {\hat{\bm \delta}}^{*2}\rangle/\langle {\hat{\bm\delta}}^*\rangle$}
\State{Solve $\hat x_{\rm{eff}}$ from \eqref{eq:dotxeff} with $\hat\beta$}
\Until{${\bm d}$ does not change}
\end{algorithmic}
\label{algo:comb_opt_round}
\end{algorithm}

\subsection{Round}
Both {\bf ZeroTopo} and {\bf TopoPlus} contain an easily unnoticed but crucial step: {\bf Round}. A rounding algorithm involves the approximation of fractional values to integer ones. In specific, Eq.\eqref{eq:missed_degree} gives a fractional $d_i$, and in our setup we consider the networks are unweighted. Therefore, we need the rounding to get an integer approximation of the degree of each node $\hat\delta_i=\delta_i^{(s)}+d_i$. It is a complicated and hard problem. In {\bf ZeroTopo} and {\bf TopoPlus} (marked with $\lfloor\cdot\rceil$), we applied the simple plain rounding algorithm because of its ease to use. However, rounding is beyond a purely numerical problem, and we should also consider its afterward impact on the predicted steady-states ${\bm z^*}$. Therefore, we design a heuristic combinatorial optimization based rounding scheme (see Algorithm~\ref{algo:comb_opt_round})  to not only ensure the accuracy in estimating individual degrees, but also minimize the error between the observed steady-states and the simulated steady-state based on the estimated degrees. The idea is to repeatedly pick a node pair $(i,j)$, then adjust their estimated missing degrees $d_i$ and $d_j$, keeping the total number of links unchanged: increase $d_i$ by 1 and decrease $d_j$ by 1 (line 8, Algorithm~\ref{algo:comb_opt_round}). The selection rule is that the two steady-states $z_i^* $ and $z_j^*$ solved from Eq.~\eqref{eq:enhanced_mf_ode} do not yet match the observed states ${\bm x^*}$ well with the current estimates $d_i$ and $d_j$. Nevertheless, after the adjustments, the steady-states ${\bm z^*}$ becomes closer to ${\bm x^*}$. Though computing seems to be very intensive, each distinct missing degree can involve in solving Eq.~\eqref{eq:enhanced_mf_ode} (line 2, Algorithm~\ref{algo:comb_opt_round}) for only once, meanwhile, the function only includes one variable. Thus, it is embarrassingly parallelizable.

\begin{table}[ht]
\centering
\resizebox{0.95\columnwidth}{!}{%
\begin{tabular}{p{19mm}p{8mm}p{8mm}p{8mm}p{8mm}p{8mm}}
\toprule
Dynamics & Net & $n$ & $m$ & $\langle {\bm\delta}\rangle$ & $\beta$\\
\midrule
\multirow{2}{*}{\textcircled{a} Ecological} & N6 & 270 &	8074 & 59.81 & 97.96\\
 & N8 & 97	& 972 & 20.04 & 32.36\\
\multirow{2}{*}{\textcircled{b} Regulatory} & ME & 2268 & 5620 & 4.96 & 40.78\\
 & TY & 662 & 1062 & 3.21 & 13.51\\
\multirow{2}{*}{\textcircled{c} Epidemic} & EM & 1133 & 5451 & 9.62 & 18.69\\
& FB & 4039 & 88234 & 43.69 & 106.57\\
\midrule
\multicolumn{6}{l}{\textcircled{a} $\dot{x}_i=B+x_i(1-\frac{x_i}{K})(\frac{x_i}{C}-1)+\sum_j\frac{x_ix_j A_{ij}}{D+E x_i+H x_j}$}\\
\multicolumn{6}{l}{\textcircled{b} $\dot{x}_i=-Bx_i^f+\sum_j A_{ij}R\frac{x_j^h}{x_j^h+1}$}\\
\multicolumn{6}{l}{\textcircled{c} $\dot{x}_i=-Bx_i+\sum_j A_{ij}R(1-x_i)x_j$}\\
\bottomrule
\end{tabular}}
\caption{Examples of real systems governed by nonlinear dynamics.  
We test our algorithms on two real networks for each dynamic. 
The notations $n$ and $m$ represent the number of vertices and edges, respectively. While $\langle {\bm\delta}\rangle$ and $\beta$ denote the average degree and resilience index respectively.}
\label{tab:networks}
\end{table}

\section{Experimental Results}\label{sec:results}
We evaluated the performance of our approaches on six real networks, governed by three dynamical equations (ecological, regulatory, and epidemic)~\cite{gao2016universal} (Table~\ref{tab:networks}).
For each network, we sample a fraction of the edges and repeat 10,000 times to get statistically significant results. On average, it takes only a few of seconds to converge. We employ the fraction of vertices with a relative error of not greater than 5\% as a proxy metric of the estimation accuracy, and the relative error is defined as ${\rm Err}[\delta_i,\hat\delta_i]=|\log(\hat\delta_i/\delta_i)|$. As reported in Table~\ref{tab:perf_degree_est}, we show the performance of each of our algorithms. Though {\bf ZeroTopo} does not take advantage of the sampled local topology information, the dynamics itself can give a decent estimate of vertex degrees even without the sampled sub-network. The {\bf TopoPlus}, either equipped with plain rounding algorithm or the designed {\bf Round} algorithm, can improve the accuracy of our estimation significantly, indicating that it is vital to utilize some available topology and rounding carefully.

\begin{table}[ht]
\centering
\resizebox{0.9\columnwidth}{!}{%
\begin{tabular}{p{4mm}cp{8mm}p{8mm}p{8mm}p{8mm}p{8mm}p{8mm}p{8mm}}
\toprule
\multirow{2}{*}{Net} & ZeroTopo & \multicolumn{3}{c}{TopoPlus} & \multicolumn{3}{c}{TopoPlus + Round}\\
\cline{2-8}
& -- & 10\% & 20\% & 30\% & 10\% & 20\% & 30\% \\
\midrule
N6 & 58.15 & 60.17 & 64.63 & 69.47 & 60.23 & 64.76 & \bf{69.65} \\
N8 & 57.73 & 61.97 & 66.46 & 71.27 & 62.74 & 67.14 & \bf{71.80}\\
ME & 86.86 & 88.02 & 90.00 & 91.82 & 89.13 & 90.91 & \bf{92.59}\\
TY & 86.16 & 87.13 & 87.81 & 88.59 & 87.76 & 88.41 & \bf{89.14}\\
EM & 59.40 & 63.20 & 67.11 & 71.30 & 65.77 & 69.75 & \bf{73.77}\\
FB & 59.54 & 64.05 & 68.82 & 73.70 & 65.36 & 70.16 & \bf{75.03}\\
\bottomrule
\end{tabular}}
\caption{Estimation accuracy on incident subgraphs with various uniform edge sampling fractions $p\in\{$10\%, 20\%, 30\%$\}$. ZeroTopo is independent of the fraction, so we identify the number as a dash.}
\label{tab:perf_degree_est}
\end{table}

\subsection{Sampling Technique}
\begin{figure}[htbp]
\centering
\includegraphics[width=0.48\textwidth]{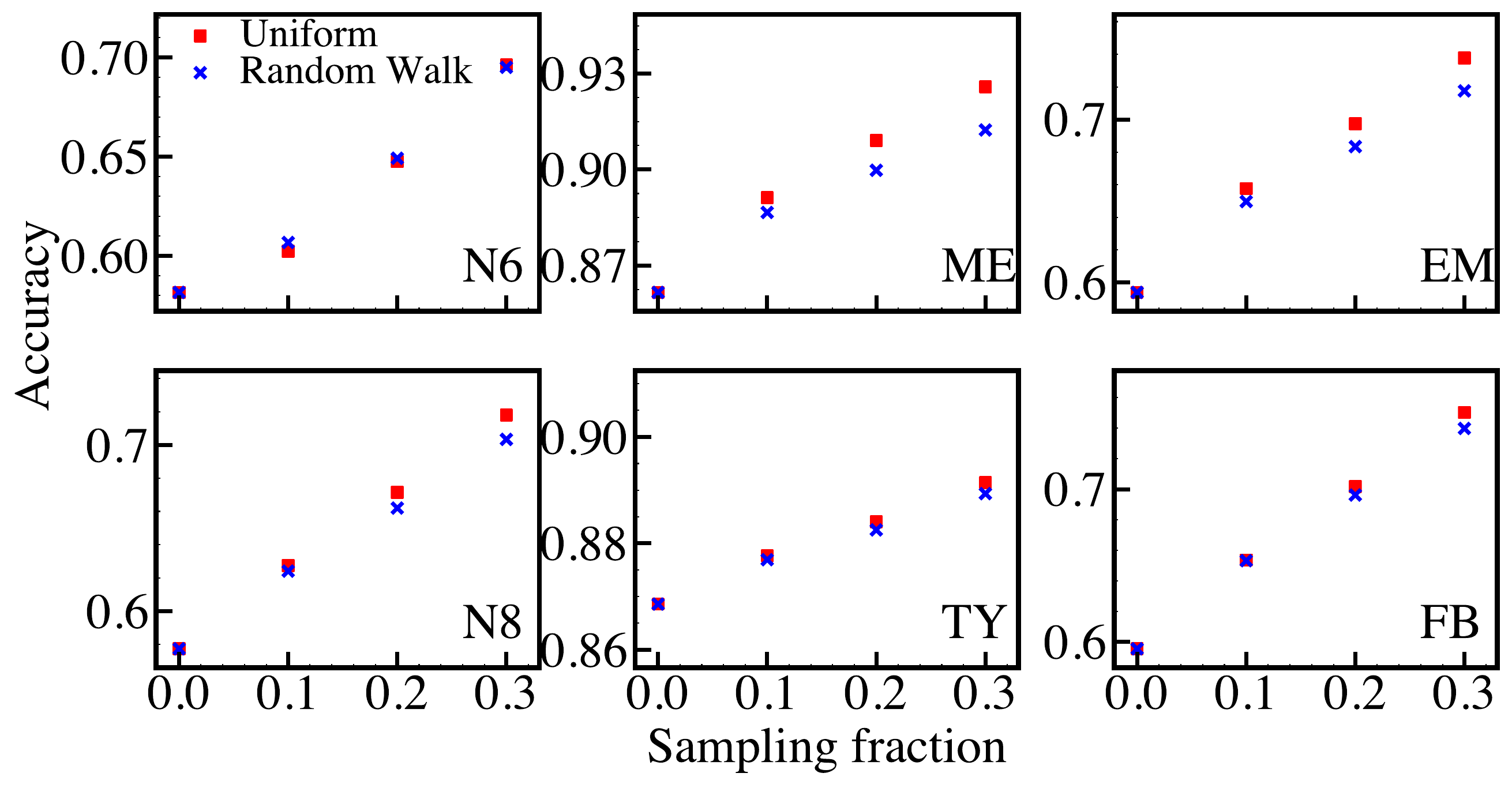}
\caption{Uniform edge sampling and random walks give similar performance. Sampling fraction $p=0$ corresponds to our approach that relies on the states only.
  \label{fig:impact_sampling}}
\end{figure}

Prior structure-based degree estimation methods~\cite{ganguly2017estimation} rely on strong assumptions about how to sample the sub-network. Specifically, these methods are designed for the induced graph from uniform sampling. In other words, it does not work for other sampling methods, for example, random walk sampling.
Our approaches do not, and they are universal, regardless of sampling methods. To demonstrate this, we compare two very different ways to sample edges: uniform sampling, and random walk sampling~\cite{hu2013survey}.
Figure~\ref{fig:impact_sampling} shows that our approaches are very robust to
the sampling scheme, as is expected.

\subsection{Measurement Noise and Model Misspecification}\label{subsec:measureerror}
\begin{figure}[ht]
\centering
\includegraphics[width=0.483\textwidth]{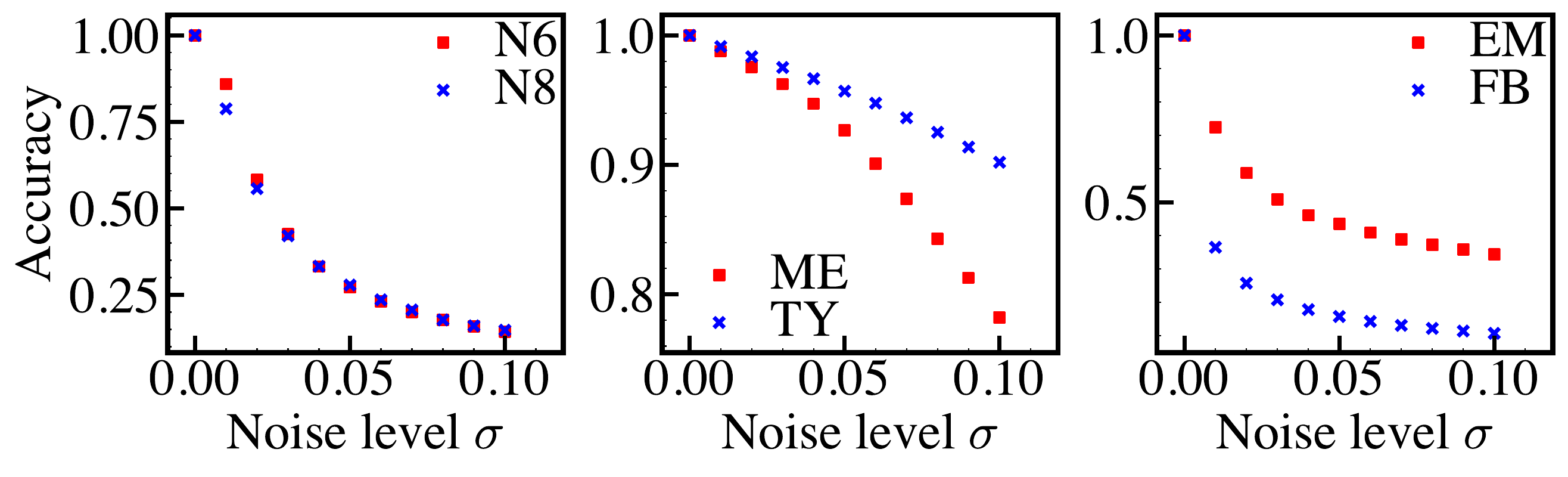}
\caption{The accuracy decreases as we increase the measurement error $\sigma$ of the observed steady-states ${\bm x}^*$ with 10\% of sampled edges. 
\label{fig:measurement_err}}
\end{figure}

Two fundamental assumptions we made are that the observed 
steady-states are measured precisely, and the true dynamics are known.
We now show that our approaches can tolerate high level of
noise added to the observed steady-states, as well as model misspecification in the form of noise added to the parameters in the dynamical system.
We add multiplicative noise $\varepsilon_x\sim N(0,\sigma)$ to the observed states, so we observed the states
${\bm x}^*(1+\varepsilon_x)\ge 0$ in simulation. Similarly, the parameters in the dynamical system are perturbed by multiplying by a factor $1+r$ for $r\in [-0.5,0.5]$.
\begin{wrapfigure}[9]{r}{.2\textwidth}
\centering
\vskip-10pt 
\includegraphics[width=.2\textwidth]{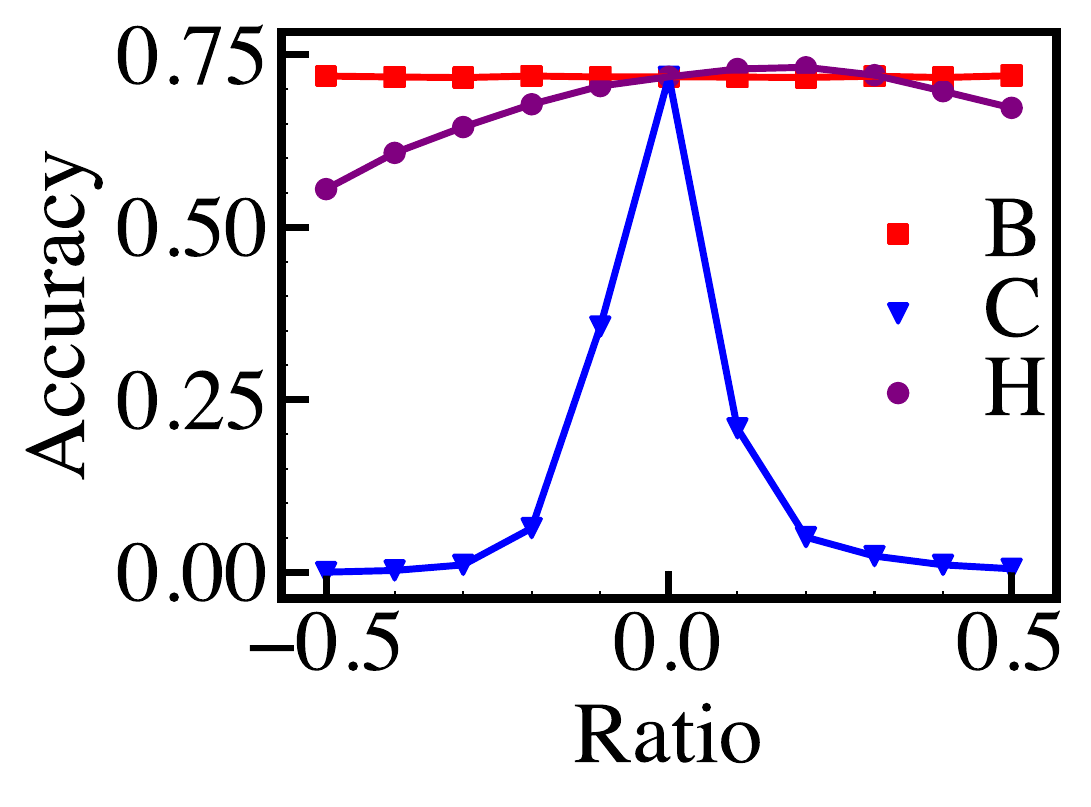}
\vskip-10pt
\caption{Impact of model misspecification.
\label{fig:model_misspecification}}
\end{wrapfigure}
Figure~\ref{fig:measurement_err} shows the robustness to
observation noise, and Figure~\ref{fig:model_misspecification}
shows the impact of errors on three parameters in the
ecological system (see Table~\ref{tab:perf_degree_est}). The parameter $B$ describes the incoming migration rate from neighboring ecosystems, $C$ controls the critical abundance of the Allee effect, and $H$ depicts the positive contribution from one vertex to another. The estimates are very sensitive to the error in the interaction parameters $C$ and $H$, but is very robust to the error in $B$.

\subsection{Local Topology and Dynamics}
We, for the first time, estimate the structure characteristics of graphs using the network dynamics.  Then, we compare the performance of our estimators and that of the structure-based estimators~\cite{ganguly2017estimation}: 
the univariate risk minimizer (\textbf{URM}), the multivariate risk minimizer (\textbf{MRM}) and the \textbf{Bayes} estimator as the baselines. Note that our estimator is regardless of the sampling methods, and the structure-based methods are only suitable for induced subgraph sampling. To ensure the fairness of the comparison, for each run, (1) we uniformly sample the same 10\% of vertices in our approaches and the structure-based estimators; (2) we only utilize the steady-state of the observed nodes in our approaches and the induced graph of the observed nodes in the structure-based estimators. We perform 10,000 runs and show the average accuracies of the baselines and our approaches in Table~\ref{tab:compare}.

\begin{table}[ht]
\centering
\resizebox{1.01\columnwidth}{!}{%
\begin{tabular}{lrrrrrrr}
\toprule
\multirow{2}{*}{Net} & \multicolumn{3}{c}{10\% \underline{\textbf{ind}}uced \underline{\textbf{sub}}graph} & 0\% ind. sub. & \multicolumn{2}{c}{10\% ind. sub.}\\
& URM & MRM & Bayes  & ZeroTopo ($\sigma$)  & TopoPlus ($\sigma$) & + Round ($\sigma$)\\
\midrule 
\multirow{2}{*}{N6} & \multirow{2}{*}{11.01} & \multirow{2}{*}{12.88} & \multirow{2}{*}{12.95} & 48.33 (0\%) & 50.57 (0\%) & 50.67 (0\%)\\
&&&&13.03 (10\%) & 13.33 (10\%) & 13.30 (10\%)\\
\midrule
\multirow{2}{*}{N8} & \multirow{2}{*}{8.28} & \multirow{2}{*}{5.30} & \multirow{2}{*}{5.10}&47.42 (0\%) & 47.93 (0\%) & 48.69 (0\%)\\
&&&&12.63 (10\%) & 12.68 (10\%) & 12.68 (10\%)\\
\midrule
\multirow{2}{*}{ME} & \multirow{2}{*}{0.63} & \multirow{2}{*}{6.27} & \multirow{2}{*}{42.79}&86.93 (0\%) & 88.70 (0\%) & 89.82 (0\%)\\
&&&&72.82 (10\%) & 73.97 (10\%) & 75.35 (10\%)\\
\midrule
\multirow{2}{*}{TY} & \multirow{2}{*}{0.54} & \multirow{2}{*}{41.74} & \multirow{2}{*}{19.96}&87.16 (0\%) & 87.73 (0\%) & 88.06 (0\%)\\
&&&&84.27 (10\%) & 84.71 (10\%) & 85.08 (10\%)\\
\midrule
\multirow{2}{*}{EM} & \multirow{2}{*}{3.74} & \multirow{2}{*}{15.00} & \multirow{2}{*}{6.61}&60.39 (0\%) & 63.95 (0\%) & 66.27 (0\%)\\
&&&&28.30 (10\%) & 29.30 (10\%) & 30.63 (10\%)\\
\midrule
\multirow{2}{*}{FB} & \multirow{2}{*}{8.23} & \multirow{2}{*}{8.81} & \multirow{2}{*}{9.00}&59.51 (0\%) & 64.03 (0\%) & 65.42 (0\%)\\
&&&&9.19 (9\%) & 9.52 (9\%) & 9.91 (9\%)\\
\bottomrule
\end{tabular}}
\caption{Estimation accuracy on induced subgraphs. The numbers in parentheses denote the additional measurement noises $\sigma$. Note that Bayes relies on the degree distribution, and we use the real one.}
\label{tab:compare}
\end{table}

We have demonstrated the promising performance of our approaches on the incident subgraphs. As we discussed in Sec.~\ref{sec:results}, they can perform well no matter how the sub-network are sampled. Therefore, they are expected to accurately estimate vertex degrees on induced subgraphs as well. Moreover, in Algorithm \ref{alg:initialization}, we assume that the states of all the nodes are known. In this comparison, we only know the states of the observed nodes. To bridge this gap, we have to estimate the global $\beta$ from the induced subgraph. Fortunately, Jiang et. al~\shortcite{jiang2020true} offer a methodology to estimate $\beta$ from incomplete networks, it can be used here. Without the states of all vertices, as shown in Table~\ref{tab:compare}, the performance of the {\bf ZeroTopo} is still outstanding compared with structure-based methods, even with 10\% noises in the measurement. It indicates that the local degrees or the neighborhood in the induced subgraph alone are not sufficient to obtain reliable estimates of individual vertex degrees. Vertex states should also be explored. There is an issue in the steady states: the measurement errors. It is ubiquitous in the real-world that the observed states are suffering from measurement noises. No measurement error (i.e., $\sigma=0$) is an ideal case, and our approaches are significantly better than the baselines. After adding a moderate amount of noise ($\sigma=10\%$ on N6, N8, ME, TY, and EM; 9\% on FB), our approaches can still beat them. The {\bf TopoPlus} and {\bf Round} provide additional improvement because of the extra topology and careful rounding.

\subsection{Application: Link Prediction}
Estimating degrees itself is an essential and exciting issue in network science. It is also useful for many other applications, such as link prediction, network reconstruction, identification of essential genes, influencer discovery, and many more. Here, we take link prediction as an example to show how our approach can enhance the current link prediction methods. The most intuitive approach to build links is computing the affinity between a node pair, e.g., (a) Adamic-Adar (AA)~\cite{adamic2003friends} $s_{u,v}=\sum_{w\in \Gamma_u\cap \Gamma_v}1/\log \delta_w$,
where $\Gamma_u$ is the neighbors of vertex $u$, and (b) Preferential Attachment (PA)~\cite{barabasi1999emergence} $s_{u,v}=\delta_u\delta_v$. Both metrics contain the degree terms. Therefore, we improve these methods by replacing the observed degrees ${\bm\delta}^{(s)}$ by the estimated degrees $\bm{\hat\delta}$ using our estimators. We proportionally scale up the size of the common neighbors $|\Gamma_u\cap \Gamma_v|$  by a factor of $\alpha_{u,v}$ according to a simple rule: $\alpha_{u,v}=\min\{\hat\delta_u/\delta_u^{(s)},\hat\delta_v/\delta_v^{(s)}\}$.

The area under the receiver-operating characteristic curve (AUC)~\cite{hanley1983method} is used to measure the prediction accuracy. AA requires a joint neighborhood, but PA does not. A small sampling fraction gives a very sparse sub-network such that joint neighborhoods are absent from $G^{(s)}$. As a result, the affinities of many node pairs are distorted. To evaluate the performance, we randomly sample $p=$1\% edges for an incident subgraph and predict the remaining 99\% missing links. As reported in Table~\ref{tab:linkpred}, the link prediction of AA is almost random. The scale-up factor can not exercise an impact on AA at all since no joint neighbors are observed.
For PA, once the individual degrees are accurately revealed, the link prediction can be of high quality as well. It can give a promising improvement on all six real networks given as low as 1\% observed edges. It demonstrates the value of our algorithm in link prediction. Note that our approach can not only improve the current link prediction methods, but it is also essential for many other inferring issues in network science.

\begin{table}[ht]
\centering
\resizebox{0.95\columnwidth}{!}{%
\begin{tabular}{p{10mm}p{9mm}p{7mm}p{7mm}p{7mm}p{7mm}p{7mm}p{7mm}}
\toprule
\multirow{2}{*}{Metrics} & \multirow{2}{*}{Status} & \multicolumn{6}{c}{Networks (Sampling $p=$1\% of edges)}\\
\cline{3-8}
& & N6 & N8 & ME & TY & EM & FB\\
\midrule
\multirow{3}{*}{AA} & {\it naive} & 50.22 & 50.06 & 50.00 &50.00 &50.02&50.27\\
&{\bf{\em revised}}& 50.22 & 50.06 & 50.00 & 50.00 & 50.02&50.27\\
& {\it truth}& 96.79 & 95.64 & 46.60 & 54.92 &86.63&99.47\\
\midrule
\multirow{3}{*}{PA} & {\it naive} & 69.07&59.90&64.13&58.00&56.18&69.44\\
 & {\bf{\em revised}}&89.29&87.11&81.18&91.28&82.33&83.78\\
 & {\it truth}&89.40&87.42&81.47&92.45&82.59&83.83\\
\bottomrule
\end{tabular}}
\caption{AUC of AA and PA, for using the observed topological information only ({\it naive}), the estimated degrees from our algorithm ({\it revised}) or the ground-truth topological information of the sampled vertices ({\it truth}). A simple scale up is made for the {\it revised} AA.}
\label{tab:linkpred}
\end{table}

\section{Discussion}

We addressed a prevalent problem.
Consider this scenario. A biologist has measured the gene expressions of all genes and collected some relationships summarized in the adjacency matrix
$A^{(s)}$. The biologist also knows how genes interact, the dynamical system \eqref{eq:dynsys_ode}. The biologist would like to identify the essential genes using the degree centrality and to make some new discoveries. Then he/she tries to use some structure-based methods to estimate the missing degrees and finds the improvement is limited. The biggest issue is that both the gene expression data and the dynamical systems are known, but he/she does not use them. In the current-state-of-the-art, no methods are available to utilize these information. The difficulty is rooted in lacking methods to theoretically solve the $n$ coupled nonlinear ODE equations with incomplete graph.
We bridge this gap by using a mean-field approximation, where the original system can be condensed to a single dimensional system with variable $x_{\rm eff}$ and topological parameter $\beta$. We provide a framework to estimate the missing degrees accurately,
which enables us to measure the essential genes.
Moreover, our approaches are robust to sampling methods and work even without knowing any topology.

There are several directions for future work.
First, we can go for a reconstruction of the
full network topology, not just vertex degrees, which needs to solve the full link-prediction problem. This is ongoing work.
Second,
the mean-field approximation is the basis of our algorithm. However, the estimation itself relies on the assumption that the network is homogeneous. Therefore, when the steady-states of the neighbors are unknown, we use $x_{\rm eff}$ as an approximation. All nodes are treated without any discrimination, although their neighborhoods may be different from each other. This assumption may not hold in real-world systems, and the estimation accuracy will be negatively affected. Another issue is the dynamical equations. In many scenarios, it is tough to calibrate the real dynamical equations. We can employ advanced machine learning approaches to extract the governing equations from an immense amount of experimental data.  

\bibliographystyle{named}
\bibliography{ref}

\end{document}